\pdfoutput=1

\documentclass{bmcart}

\RequirePackage{hyperref}
\usepackage[utf8]{inputenc} 

\usepackage{graphicx}
\usepackage{comment}
\usepackage{amsmath}
\usepackage{amssymb}
\usepackage{algo}
\usepackage{cite}

\usepackage[lofdepth,lotdepth,font={footnotesize,sf},caption=false]{subfig}

\newcommand{\examine}{\textsf{eXamine}\xspace}

\graphicspath{{images/}}

\startlocaldefs
\endlocaldefs

\begin{document}

\begin{frontmatter}

\begin{fmbox}
\dochead{Software}


\title{eXamine: a Cytoscape app for exploring annotated modules in networks}


\author[
   addressref={aff1},                   
   noteref={n1},                        
   email={k.dinkla@tue.nl}              
]{\inits{K}\fnm{Kasper} \snm{Dinkla}}
\author[
   addressref={aff2,aff3},
   noteref={n1},                        
   email={m.el-kebir@cwi.nl}
]{\inits{M}\fnm{Mohammed} \snm{El-Kebir}}
\author[
   addressref={aff2,aff3},
   email={bucur@cwi.nl}
]{\inits{C-I}\fnm{Cristina-Iulia} \snm{Bucur}}
\author[
   addressref={aff3},
   email={m.siderius@vu.nl}
]{\inits{M}\fnm{Marco} \snm{Siderius}}
\author[
   addressref={aff3},
   email={mj.smit@vu.nl}
]{\inits{MJ}\fnm{Martine J.} \snm{Smit}}
\author[
   addressref={aff1},                   
   corref={aff1},
   email={m.a.westenberg@tue.nl}              
]{\inits{MA}\fnm{Michel A.} \snm{Westenberg}}
\author[
   addressref={aff2,aff3},
   email={gunnar.klau@cwi.nl}
]{\inits{GW}\fnm{Gunnar W.} \snm{Klau}}

\address[id=aff1]{
  \orgname{Eindhoven University of Technology}, 
  \street{Den Dolech 2},                     %
  \postcode{5600 MB},                                
  \city{Eindhoven},                              
  \cny{The Netherlands}                                    
}
\address[id=aff2]{%
  \orgname{Life Sciences, Centrum Wiskunde \& Informatica (CWI)},
  \street{Science Park 123},
  \postcode{1098 XG},
  \city{Amsterdam},
  \cny{The Netherlands}
}
\address[id=aff3]{%
  \orgname{VU University Amsterdam},
  \street{De Boelelaan 1105},
  \postcode{1081 HV},
  \city{Amsterdam},
  \cny{The Netherlands}
}


\begin{artnotes}
\note[id=n1]{Equal contributor} 
\end{artnotes}

\end{fmbox}


\begin{abstractbox}

\begin{abstract} 
\parttitle{Background} Biological networks have growing importance for the interpretation of
high-throughput ``omics'' data. 
Statistical and combinatorial
methods allow to obtain mechanistic insights through the extraction of smaller subnetwork modules. Further enrichment analyses provide set-based annotations of these modules.
\parttitle{Results} 
We present \examine, a set-oriented visual analysis approach for annotated
modules that displays set membership as contours on top of a
node-link layout. Our approach extends upon Self Organizing Maps to simultaneously lay out nodes, links, and set
contours. 
\parttitle{Conclusions} We implemented \examine as a freely available Cytoscape app. Using \examine we study a module that is activated by the virally-encoded G-protein coupled receptor US28 and formulate a novel
hypothesis about its functioning.
\end{abstract}






\end{abstractbox}
%

\end{frontmatter}




\section*{Background}
\label{s:introduction}

\begin{figure}[hbtp]
  \centering
  \includegraphics[width=0.9\textwidth]{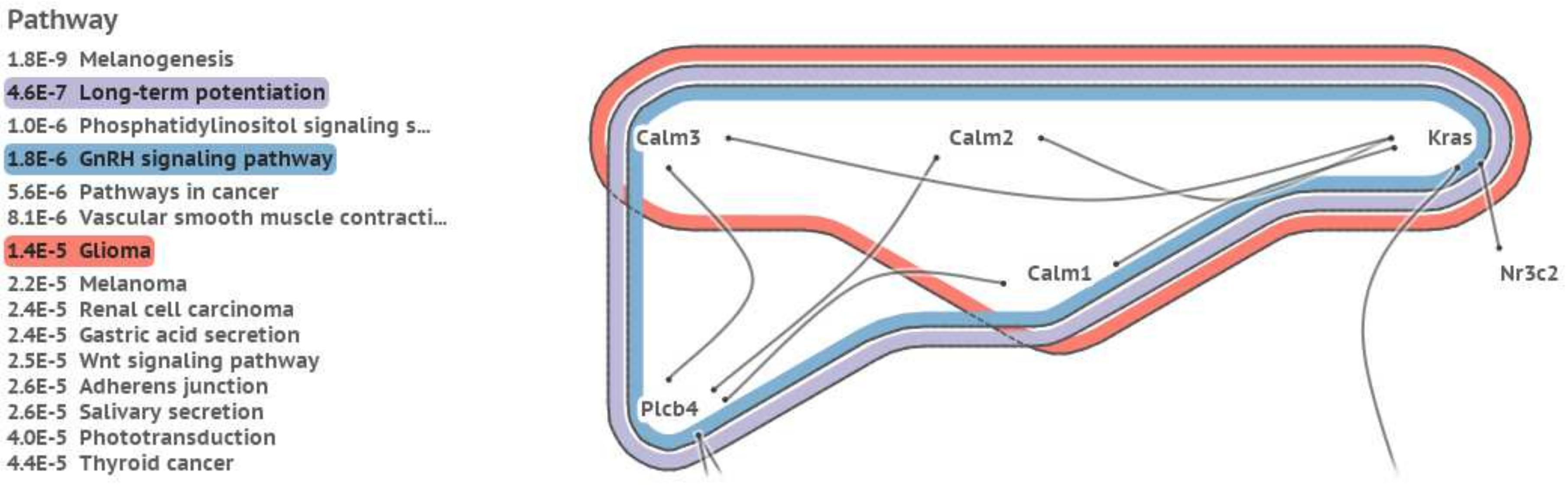}
  \caption{\csentence{Visualization of an annotated module.} Interacting
    proteins with a selection of three subsets, corresponding to
    overrepresented KEGG pathways. The visualization consists of a
    combination of a node-link diagram and an Euler diagram.}
  \label{f:teaser}
\end{figure}

Traditionally, computational approaches to analyze high-throughput ``omics'' data
have been \emph{gene-centric} and typically result in ranked lists of
differentially expressed genes~
\cite{Alizadeh:2000ko,Golub:1999un,vandeVijver:2002dt}.
Later, gene-centric approaches have been
complemented by \emph{pathway-}~\cite{Tarca:2009hf,Vaske:2010is} and
\emph{network-based} methods~\cite{Ideker:2002bd,Dittrich08}. While
pathway-based approaches identify overrepresented pathways from databases such
as the Kyoto Encyclopedia of Genes and Genomes (KEGG)~\cite{Kanehisa:2000jn},
network-based approaches yield small, \emph{de novo} subnetwork modules 
that may span several pathways taking their crosstalk into
account~\cite{Mitra:2013em}. To assess the significance of such a module, a
subsequent overrepresentation analysis is performed to identify enriched
categories originating from ontologies such as the Gene Ontology
(GO)~\cite{Ashburner:2000ja} or enriched pathways from
KEGG~\cite{Kanehisa:2000jn}. This results in an \emph{annotated
  module}, which, in addition to its network structure, captures the
enrichment information using a \emph{set system}, that is, a family of 
possibly overlapping sets of nodes---see Figure~\ref{f:teaser} for an
illustration. The interpretation of annotated modules can
be facilitated using visualization techniques. 

There are many tools for interpreting and exploring biological
networks~\cite{gehlenborg10}, including the popular open source
platforms Cytoscape \cite{smooth11} and PathVisio
\cite{vanIersel:2008ex}. However, they currently provide only limited
capability to visualize annotated modules. 
PathVisio is a pathway analysis approach, where sets are restricted to subsets of
static, pre-defined individual pathways and set membership is shown by node colors.
Cytoscape's group
attributes layout can be used to visualize partitions by showing
disjoint parts in separate circles, but does not support
overlapping sets. The Venn and Euler diagram app\footnote{\url{http://apps.cytoscape.org/apps/vennandeulerdiagrams}}
for Cytoscape does support overlapping sets, but their number is limited
to four. In this app, network and sets are visualized separately: set membership is conveyed by selecting a set and
its corresponding nodes are highlighted in Cytoscape's network
view (see Figs.~\ref{f:venn} and \ref{f:euler}). The RBVI 
collection of plugins\footnote{\url{http://www.rbvi.ucsf.edu/cytoscape/groups/}} facilitates creation
and editing of Cytoscape groups, and provides a group viewer that
relies on aggregation of groups into meta-nodes. These meta-nodes can be
visualized as standard nodes, as nodes containing embedded networks,
or as charts. This approach, however, does not allow for visualization
of overlapping sets.

In the information visualization field, \emph{Euler diagrams} are used
for an intuitive visualization of set systems~\cite{bertault01, simonetto09, riche10}, where items belonging
to the same set are denoted by contours. Variants of these approaches visualize sets over items with
predefined positions, e.g., over a given node-link visualization of a
network. These methods range from connecting these items by
simple lines (LineSets)~\cite{alper11}, via colored shapes that are routed along
the items (Kelp Diagrams)~\cite{dinkla12kelp} and contours around the
items (BubbleSets, see Fig.~\ref{f:bubble_sets})~\cite{collins09,smith12} to hybrid
approaches (KelpFusion)~\cite{meulemans13}. 
Visualizing an annotated module, however, requires an integrated
layout of both its network and set system topologies, which is not
possible with these existing approaches. Euler diagram methods focus on the layout of the set
relations at the expense of the network topology. Likewise, laying out
the network first before superimposing set
relations will emphasize network topology to the detriment of the set system.
There exist some techniques that provide such integrated
layouts~\cite{sugiyama91,shneiderman06,barsky08}, but they 
assume constraints on the network and set system
topologies, e.g., strict partitions and no overlapping sets, and are therefore
not applicable to our problem.

In this paper, we present a visual analysis approach designed around
the typical characteristics of annotated network modules: (i) small
and sparse network topology, with nodes and edges that number in the
dozens; (ii) large set system, where sets outnumber edges and vary in
size, overlap, and distribution of set member nodes across the
network.  Our approach displays sets as contours on top of a node-link
layout, however, in contrast to previous work mentioned above, our
approach places emphasis on the interactive exploration of a large set
system together with an underlying network topology, see
Fig.~\ref{f:teaser}.  We do this in a unified way using \emph{Self
  Organizing Maps} (SOMs)---also known as Kohonen maps. SOMs map
high-dimensional data to discretized low-dimensional
space~\cite{kohonen90} and have successfully been applied to lay out
networks~\cite{meyer98}.  We have implemented this approach as a
generic Cytoscape app \examine, such that it can be applied to
annotated network modules as they appear for various domain-specific
networks.

 




We apply \examine to study an annotated module that is activated by
the virally-encoded G-protein coupled receptor US28. We obtained the
module by analyzing microarray expression data of US28 vs.\ mock
transfected murine cell lines within the context of
a network consisting of murine signaling and metabolic pathways from KEGG.  The annotation
consists of enriched GO-terms and KEGG pathways as well as gene expression fold
changes. Using \examine, we formulate a novel hypothesis about
deregulated signaling of $\beta$-catenin by the viral receptor protein
US28.

\section*{Method and Implementation}

Visualizing an annotated module amounts to visualizing a
\emph{hypergraph} consisting of binary
edges, representing the network structure, and
$n$-ary edges, representing the set system on the nodes. As opposed to combining
multiple existing techniques---e.g., a force simulation to position the nodes
according to the binary edges~\cite{fruchterman91},
a node overlap removal algorithm to keep nodes identifiable~\cite{dwyer06},
and subsequent construction of a density field to derive contours for
$n$-ary edges~\cite{collins09}---our approach uses a single technique for
both visualization tasks. To this end, we assign a bit vector $\mathbf{t} =
(t_1, t_2, \ldots,t_M)$ to every
node $t \in V$ (the set of nodes in the annotated module) that encodes its membership in binary and $n$-ary edges $S_1,
S_2,\ldots,S_M$. That is, $t_i = 1$ if $t \in S_i$ and $t_i = 0$ if $t \not \in S_i$.

To make this representation more concrete, consider the annotated
module shown in Fig.~\ref{f:teaser}. The nodes are represented as the
set $V = \{$Calm1, Calm2, Calm3, Kras, Nr3c2, Plcb5$\}$. There are
seven sets representing the edges and three sets representing pathway
memberships. The edge sets are
$S_1 = \{v_1,v_4\}$, $S_2=\{v_1,v_6\}$, $S_3=\{v_2,v_4\}$, $S_4=\{v_2,v_6\}$,
$S_5=\{v_3,v_4\}$, $S_6=\{v_3,v_6\}$, and $S_7=\{v_4, v_5\}$. Note
that nodes $v_4$ (Kras) and $v_6$ (Plcb4) have some additional
outgoing edges, but their targets are not visible in the
image. Therefore, we ignore these edges in this example. The pathway
memberships are 
the
Glioma set $S_8 = \{v_1, v_2, v_3, v_4 \}$, the  Long-term
potentiation set $S_9 = \{v_1, v_2,
v_3, v_4, v_6\}$, and the GnRH
signaling pathway set $S_{10} = \{ v_1, v_2,
v_3, v_4, v_6\}$. Now, for example, node $v_5$  gets assigned the bit vector
$\mathbf{t}_{v_5} = (0,0,0,0,0,0,1,0,0,0)$, and node $v_6$ the bit vector
$\mathbf{t}_{v_6} = (0,1,0,1,0,1,0,0,1,1)$. 

This high-dimensional representation is then used to lay out the nodes
without overlap, the binary edges as curves and the $n$-ary edges as contours.



\subsection*{Extension to Self Organizing Maps}
\label{ss:som}
\begin{figure}[hbtp]
  \centering
  \subfloat[]{\includegraphics[width=4cm]{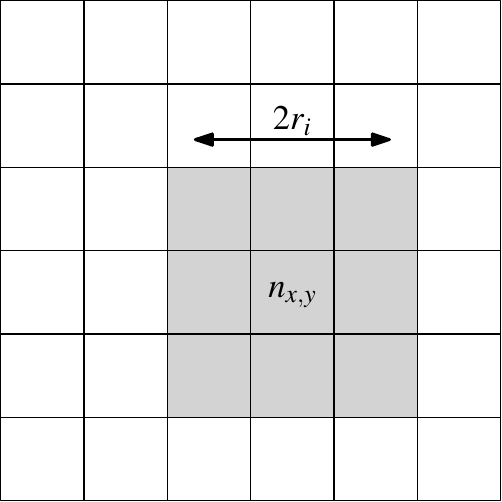}}\hfill
  \subfloat[]{\includegraphics[width=4cm]{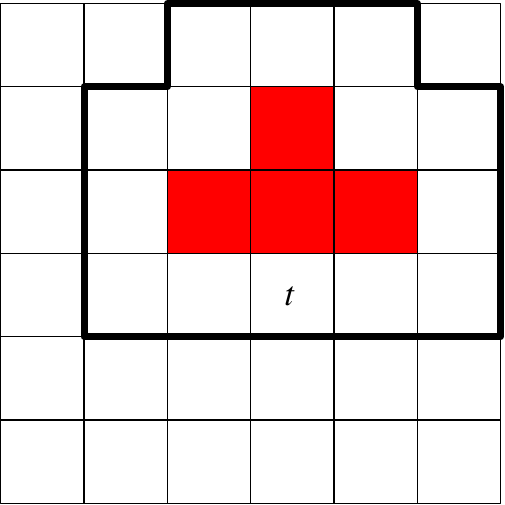}}

  \caption{\csentence{Training neuron $n_{x,y}$.}
(a) The neighborhood within range $r_i$ is trained (colored gray).
(b) Certain tiles are already reserved (colored red) in the
    \emph{RSOM} algorithm, item $t$ therefore trickles outwards
    to the best matching free spots (outlined).}
  \label{f:som_training}
\end{figure}

Self Organizing Maps (\emph{SOMs}), introduced by Kohonen~\cite{kohonen90}, are
an artificial neural network that are used to map high-dimensional data items
to discretized low dimension. In a visualization setting SOMs are used to
cluster similar items together in a 2D embedding, creating a landscape
of items based on their features~\cite{vesanto99, maccallum11}. Typical SOMs consist of a square
grid of size $N \times N$ with a neuron $n_{x,y} \in [0..1]^{M}$ at every grid
cell. A neuron $n_{x,y}$ is a bit vector of size $M$ whose dimension matches
the data items' dimensions. In our case the data items $\mathbb{T}$ correspond
to the set of nodes $V$ in the annotated module. The training algorithm applies unsupervised
reinforcement learning in an iterative fashion: at every iteration $i \in
\{1,\ldots,I\}$ all data items $t \in
\mathbb{T}$ are considered and the neuron that matches $t$ most closely is
determined using a distance function such as the Euclidean or Manhattan norm.
This neuron and its neighboring neurons within radius $r_i$ are updated
to match $t$ even more closely by setting their respective vectors $q$ to $q +
\alpha_i (t - q)$---see Fig.~\ref{f:som_training}(a).
In early iterations $i$, the trained neighborhoods are large with
$r_{i}$ close to the grid size $N$ and the training strength
$\alpha_{i}$ close to 1. The parameters $r_i$ and $\alpha_i$
decrease monotonically with increasing $i$. As such, items that differ strongly
will distribute across the map to establish their own regions in the grid at
early stages. Items with smaller differences are separated along the grid at a
more local level as the training iterations progress.

\medskip\noindent\textbf{Reservation based training.}
The drawback of standard SOMs is that similar items may end up at the same grid
position. Usually this is solved by showing aggregate depictions of items, but
in our case we want to have separate depictions without overlap. Therefore,
each item has to map to a unique grid position. We achieve this by altering the
training algorithm:
\begin{algorithm}
{RSOM}{$(\mathbb{T})$}
\qfor $i$ \qlet\ $1$ \qto $I$\\
  \qdo Initialize copy $\mathbb{U}$ of $\mathbb{T}$ and clear neuron reservations.\\
       \qwhile $\mathbb{U}$ contains items\\
       \qdo Draw and remove item $t$ from $\mathbb{U}$.\\
            Find unreserved neuron $n_{x,y}$ with smallest distance $d(t,
            n_{x,y})$.\\
            Reserve $n_{x,y}$ for $t$.\\
            \qfor any neuron $q$ within range $r_i$ from $(x,y)$\\
            \qdo $q$ \qlet\ $q + \alpha_i (t - q)$
            \qendfor
    \qendfor
  \qendfor
\end{algorithm}
Now items are actively assigned to a unique neuron after every training iteration
because once a neuron is reserved by an item, subsequent items will ignore it. This
causes a flooding effect where similar items end up in the same area of the grid
and trickle outwards as the area becomes more crowded---see Fig.~\ref{f:som_training}(b).

\medskip\noindent\textbf{Configuration.}
As the distance function $d$ we use the metric distance form of cosine
similarity, i.e.~$d(q, p) = \cos^{-1}((q \cdot p)(|q| |p|))
\pi^{-1}$. For our data items this measure performs better than the Euclidean and
Manhattan norms. We train the SOM with a learning strength and neighborhood range
that decrease linearly with increasing iteration $i$, i.e. $\alpha_i = 0.01 \cdot (1 - i/I)$ and $r_i = \lfloor
(1- i/I) \cdot N \rfloor$. As for the number of neurons and iterations we use 
$N = {2 |\mathbb{T}|}$ and $I = 10^5/|\mathbb{T}|$, respectively.
Instead of using a square tiling of the neurons, we use a hexagonal topology as
it results in improved set contour aesthetics.

\medskip\noindent\textbf{Layout preservation.}
Whenever the user selects or deselects a set, a new layout has to be computed.
In order to preserve the user's mental map, the new layout should change
only slightly in comparison to the old layout.
We achieve this by creating a new SOM and using the old layout as the initial
configuration of the neurons, i.e., an item that
was positioned at $n_{x,y}$ in the old SOM is placed at $n_{x,y}$ in the new SOM
and its neighborhood is trained according to the new bit vector of the item.
We ensure that the new SOM retains the starting configuration as much as
possible by starting the training factor $\alpha_i$ at value 0.01.
Naturally, this imposes a trade-off between layout quality and conservation.

\begin{figure}[hbtp]
  \centering
  \subfloat[]{\includegraphics[width=4cm]{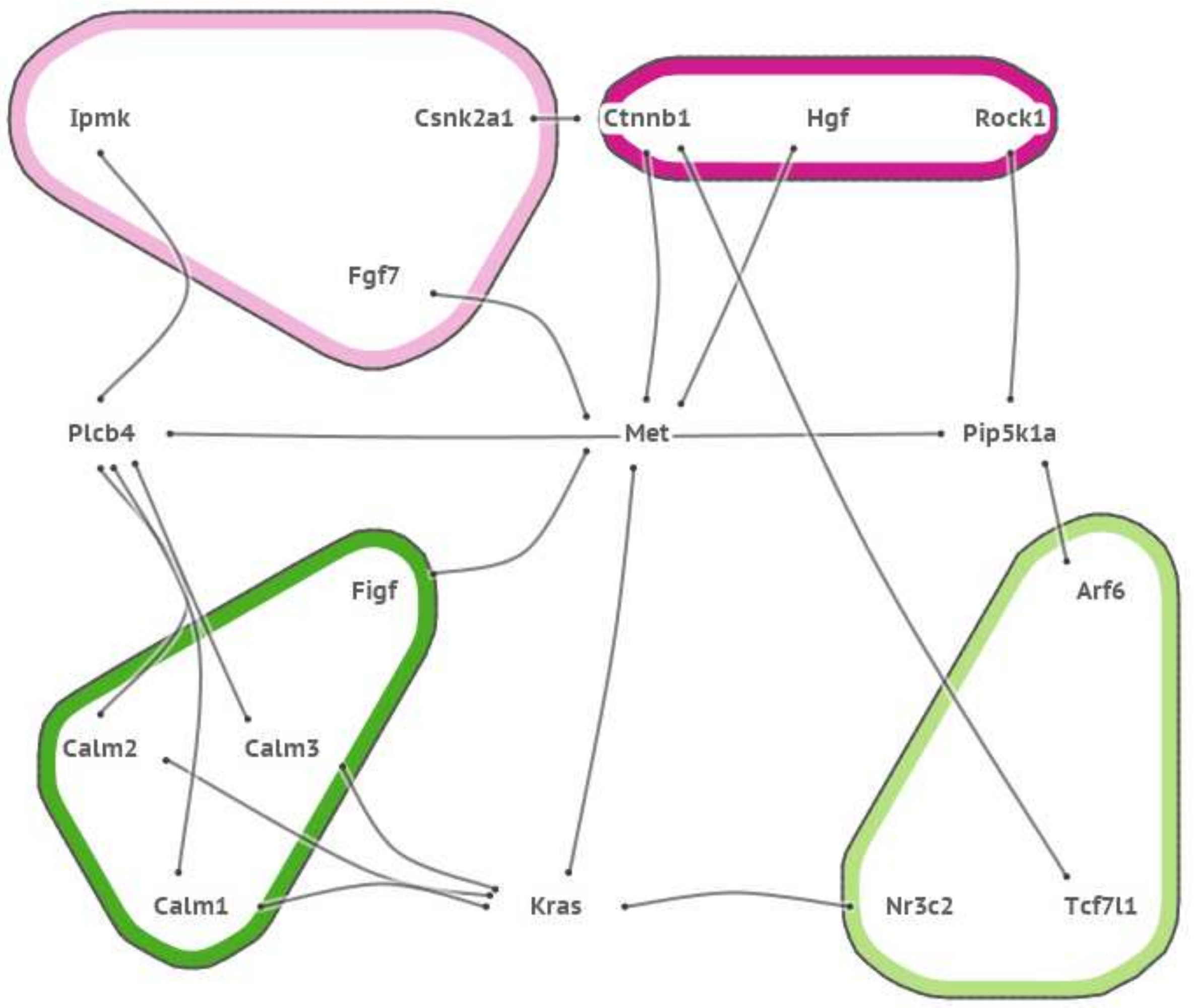}}\hfill
  \subfloat[]{\includegraphics[width=4cm]{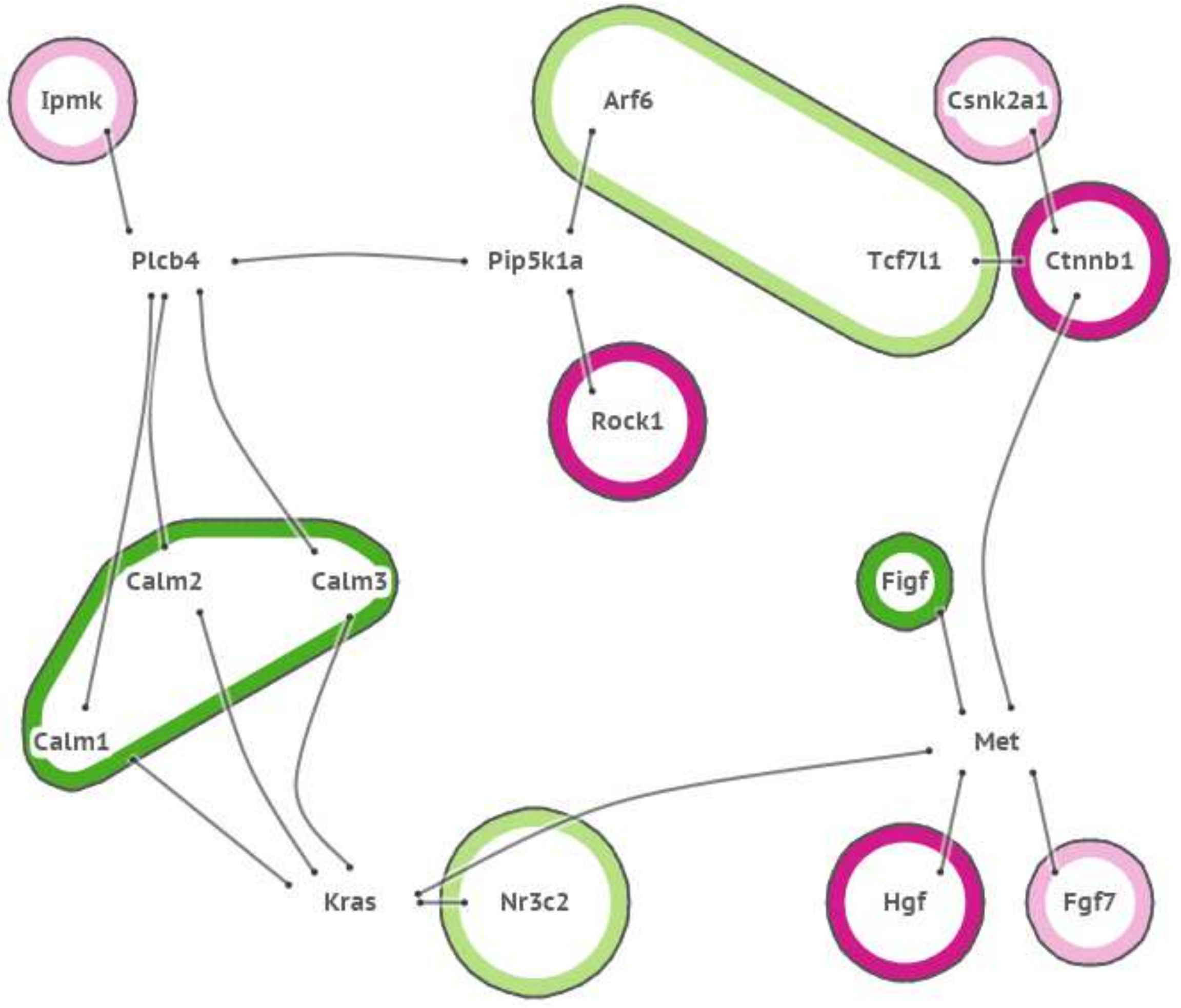}}

  \caption{\csentence{Changing the dominance of sets.}
(a) Sets are highly dominant, drawing proteins of the same sets together and bundling their interactions.
(b) Sets have no dominance such that the network topology fully defines the layout.}
  \label{f:dominance}
\end{figure}

\medskip\noindent\textbf{Set dominance.}
We allow the user to make a certain set more dominant in the layout
by having the training algorithm place the items of that set closer to each other than the items of other
sets. We do this by weighing the components of the item bit vectors:
every $S_i$ is given a weight $w_i$ where initially $w_i = 1$. The bit vectors
are subsequently augmented to incorporate these weights: $t_i = w_i$ if $t \in
S_i$ and $t_i = 0$ if $t \not\in S_i$. The bit vector component of $S_i$
will therefore play a more prominent role in distance metric $d$ when the user
increases $w_i$---see Fig.~\ref{f:dominance}.

\begin{figure}[hbtp]
  \centering
  \subfloat[]{\includegraphics[width=4cm]{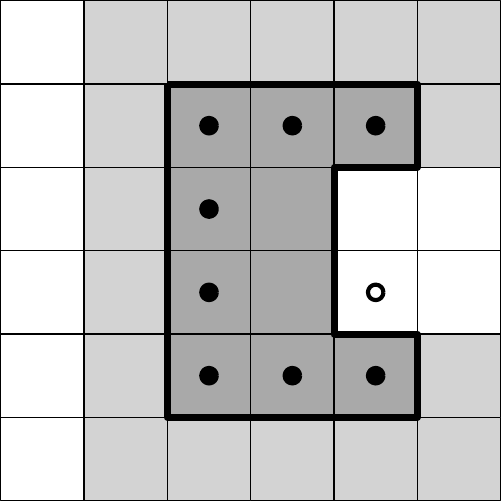}}\hfill
  \subfloat[]{\includegraphics[width=4cm]{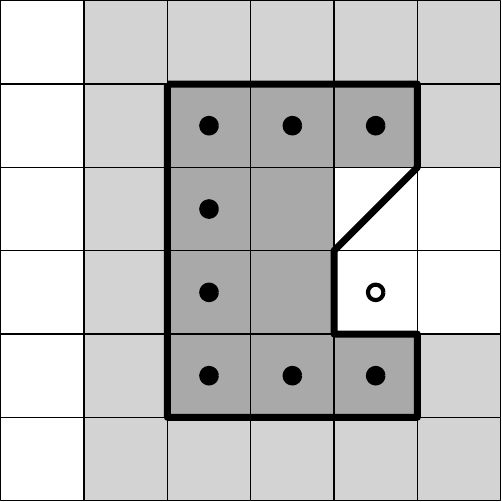}}

  \caption{\csentence{Derivation of contours for set $S_i$.} The darkness
  of a tile represents the value of the neurons' $i$-th component,
  the thick black line is the contour,
  dots represent items that are in $S_i$, and white dots are items that are not in $S_i$.
  (a) Contour that results from the union of tiles with a value
  above a certain threshold.
  (b) Refined contour with shortcuts across free tiles.}
  \label{f:contours}
\end{figure}

\medskip\noindent\textbf{Contours.}
We use the SOM's neuron grid to define the contours representing the active set
system. Let $S_i$ be an active set. The corresponding $i$-th components of 
the neurons define a scalar field that essentially is a fuzzy membership
landscape for $S_i$. This field is similar to the density field used in Bubble
Sets~\cite{collins09}. Now, the inclusion of the grid tile
of neuron $n$ in the contour body is determined by imposing a threshold,
of for example $\frac{1}{2}$, on the $i$-th component (see Fig.~\ref{f:contours}(a)).
The contour can then be tightened to reduce sharp corners
by including parts of tiles that are free of items, as illustrated in Fig.~\ref{f:contours}(b).

\begin{figure}[hbtp]
  \centering
  \subfloat[]{\includegraphics[width=4cm]{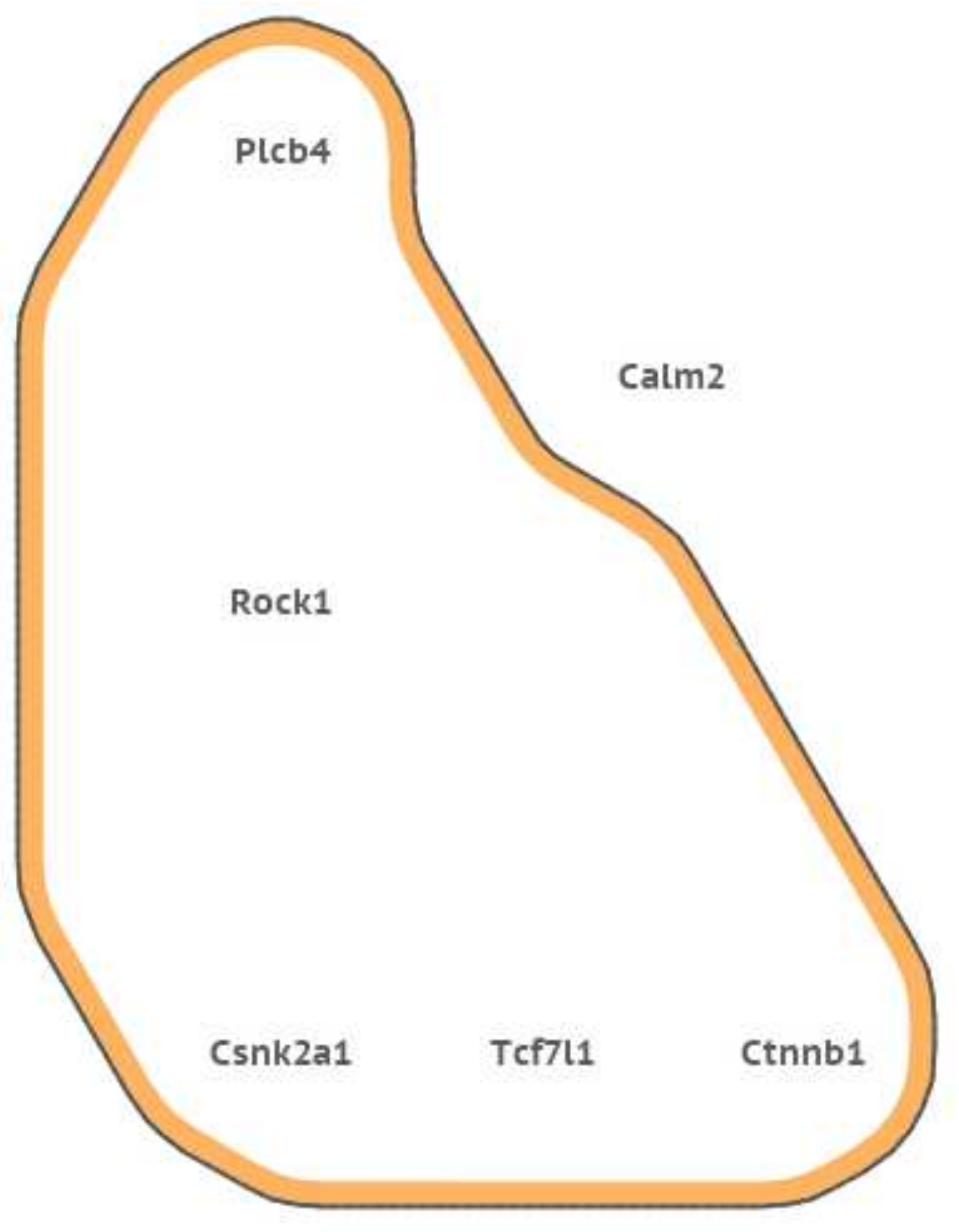}}\hfill
  \subfloat[]{\includegraphics[width=7cm]{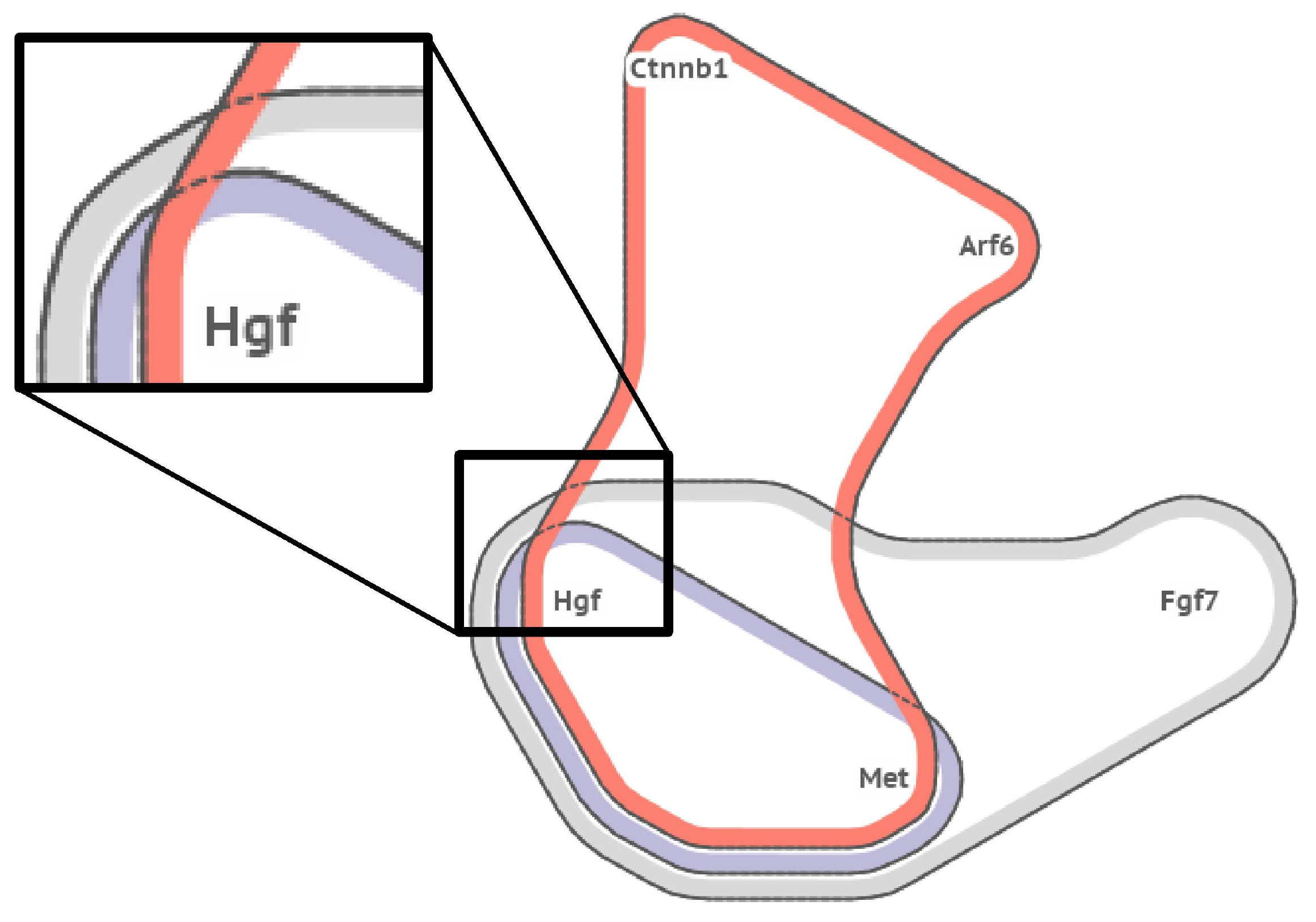}}

  \caption{\csentence{Geometric refinement of set contours after initial layout.}
(a) Corners are smoothened by dilation and erosion operations, and contours are
given a thick and colored internal ribbon.
(b) Unique erosion levels create distance between contour outlines, and
contour overlap is emphasized by dashed lines.}
  \label{f:contours_details}
\end{figure}

After establishing the layout of the contours, we apply geometric post
processing \cite{dinkla12kelp} to improve aesthetics.
We round the sharp corners of the initial layout by a
dilation of $r$, erosion of $2r$, and subsequent dilation of $r$---see Fig.~\ref{f:contours}. 
Here \textit{dilate} and \textit{erode} are equivalent to \textit{Minkowski sum} and
\textit{Minkowski subtraction} operators with a circle of radius $r$~\cite{berg08}. In addition,
we nest the contours by applying different levels of erosion, enforcing
a certain distance between them. We obtain the thick colored ribbons in Fig.~\ref{f:contours_details} 
by taking the body $b$ of a contour, eroding it to get a smaller body $b_e$, and
taking the symmetric difference $b - b_e$ of $b$ and $b_e$---effectively cutting
$b_e$ out of $b$.
We bound radius $r$, i.e., the extent of the erosions and dilations, by a fraction
of the grid's tile size such that: no items end up being covered by a smoothened contour of
$S_i$ while not being part of $S_i$, and no contour sections of $S_i$ are smoothened out while
covering items that are part of $S_i$.

We draw the contours in the order of their nesting caused by their
different erosion levels; the largest contour is drawn first and the
smallest contour last. We assign the contour ribbons unique colors per set
and draw them fully opaque to prevent any confusion caused by blended colors. We
mitigate occlusion by limiting the width of the ribbons. Moreover, we draw the contours
a second time as dashed lines such that occluded contour sections can be
inferred---see Fig.~\ref{f:contours_details}(b).

\begin{figure}[hbtp]
  \centering
  \subfloat[]{\includegraphics[width=4cm]{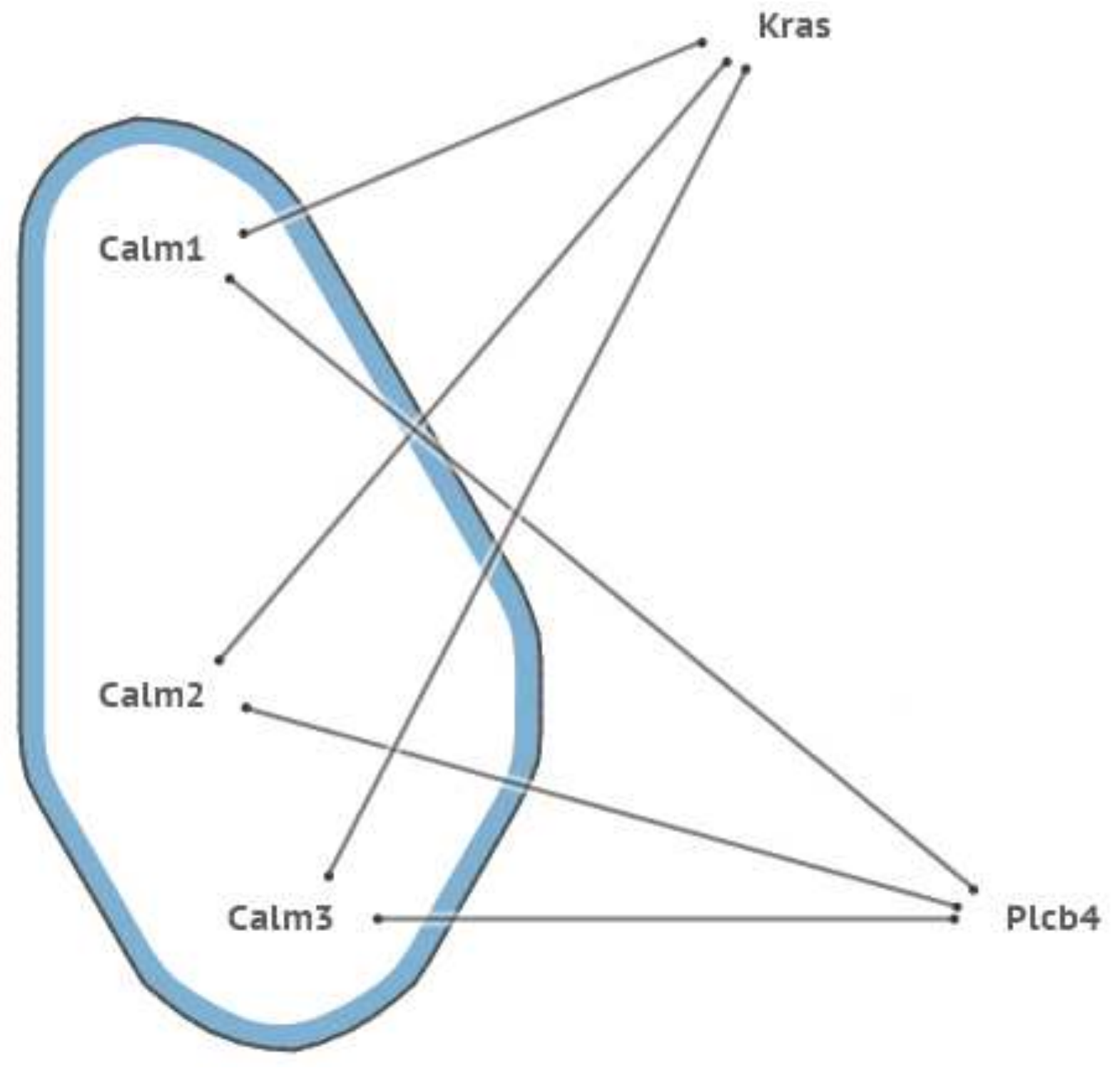}}\hfill
  \subfloat[]{\includegraphics[width=4cm]{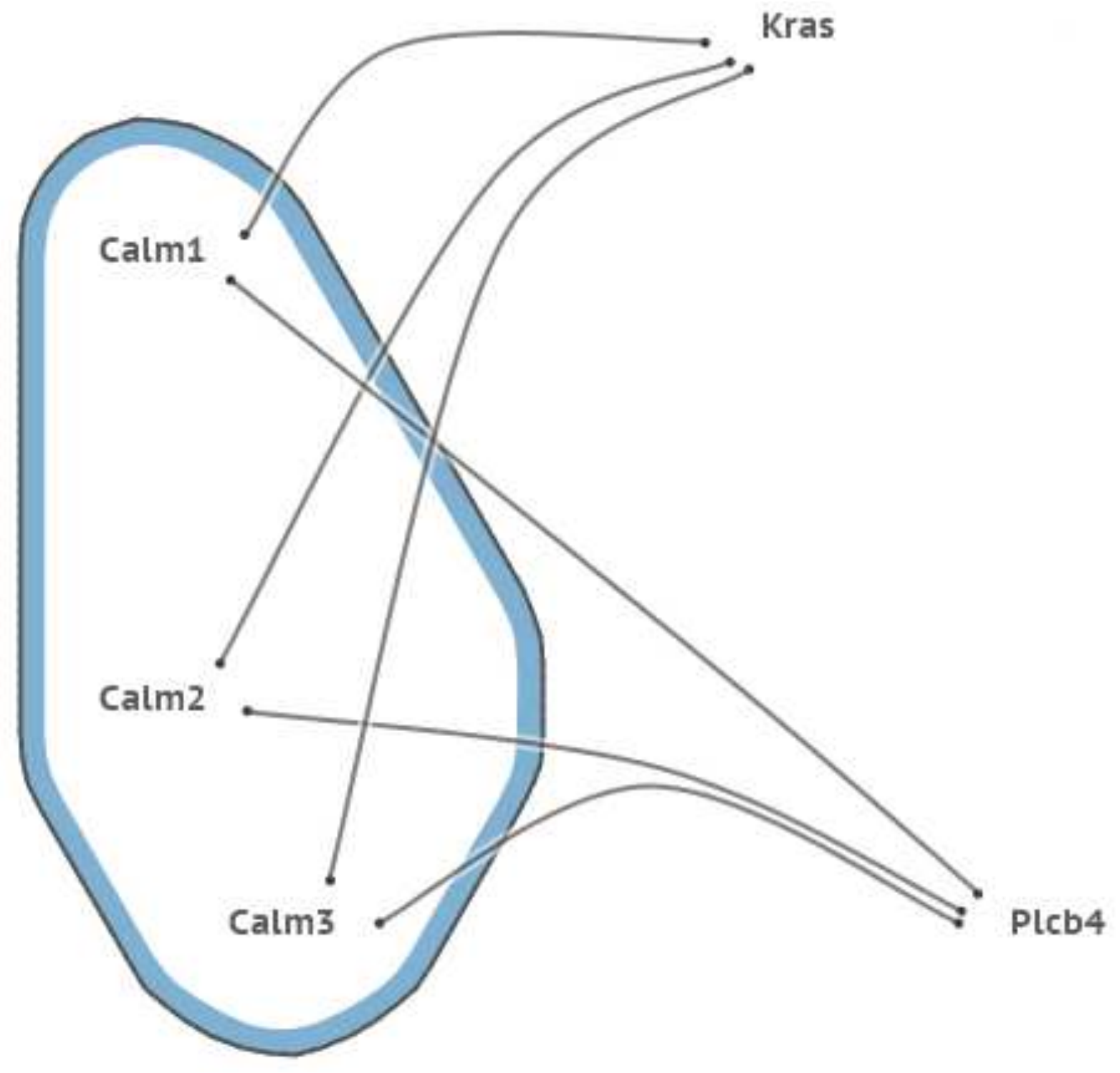}}

  \caption{\csentence{Link layout variants.}
(a) Straight lines that connect items.
(b) Splines that bundle when connecting similar items.}
  \label{f:links}
\end{figure}

\medskip\noindent\textbf{Links.}
Instead of drawing edges as straight lines (see Fig.~\ref{f:links}(a)),
we use the SOM's neuron grid to bundle links.
We do this by drawing a spline between two nodes along three control points~\cite{holten09}.
These control points are derived by first linearly interpolating between the two
nodes in the high-dimensional space that describe their set memberships, and then
projecting the high-dimensional control points to the 2D space of the neuron grid.
This projection is based on inverse distance weighting of the neuron's positions, using
Shepard's method~\cite{Shepard:1968} with distance function $d$ for the weighting.
This gives those neurons closest to a control point, according to $d$, most
influence on the control point's 2D position, resulting in
a continuous mapping from the high-dimensional space of
the items to the 2D layout space. Hence, the link $(s,t)$ is guided by the neurons as it moves from node
$s$ to $t$.
No routing is performed to prevent node and link intersections, nor does
the described bundling approach suppress these situations.
Instead, we avoid ambiguous situations by terminating links at ample
distance from the nodes that they connect and marking their terminations with dots.

\subsection*{Implementation}
\label{s:tool}
\begin{figure}[hbtp]
  \centering
  \subfloat[]{\includegraphics[width=0.9\textwidth]{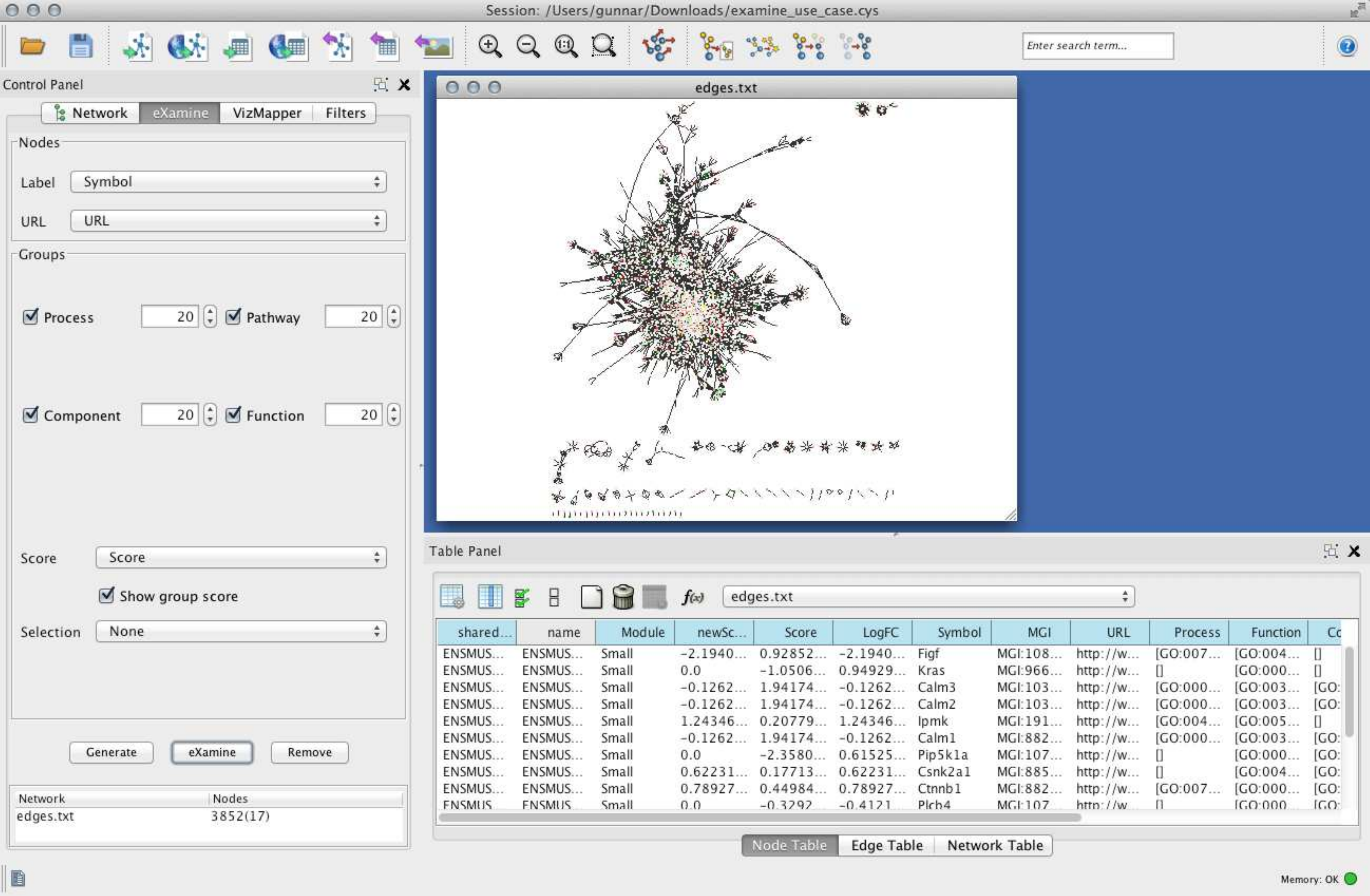}}
  \\
  \subfloat[]{\includegraphics[width=0.9\textwidth]{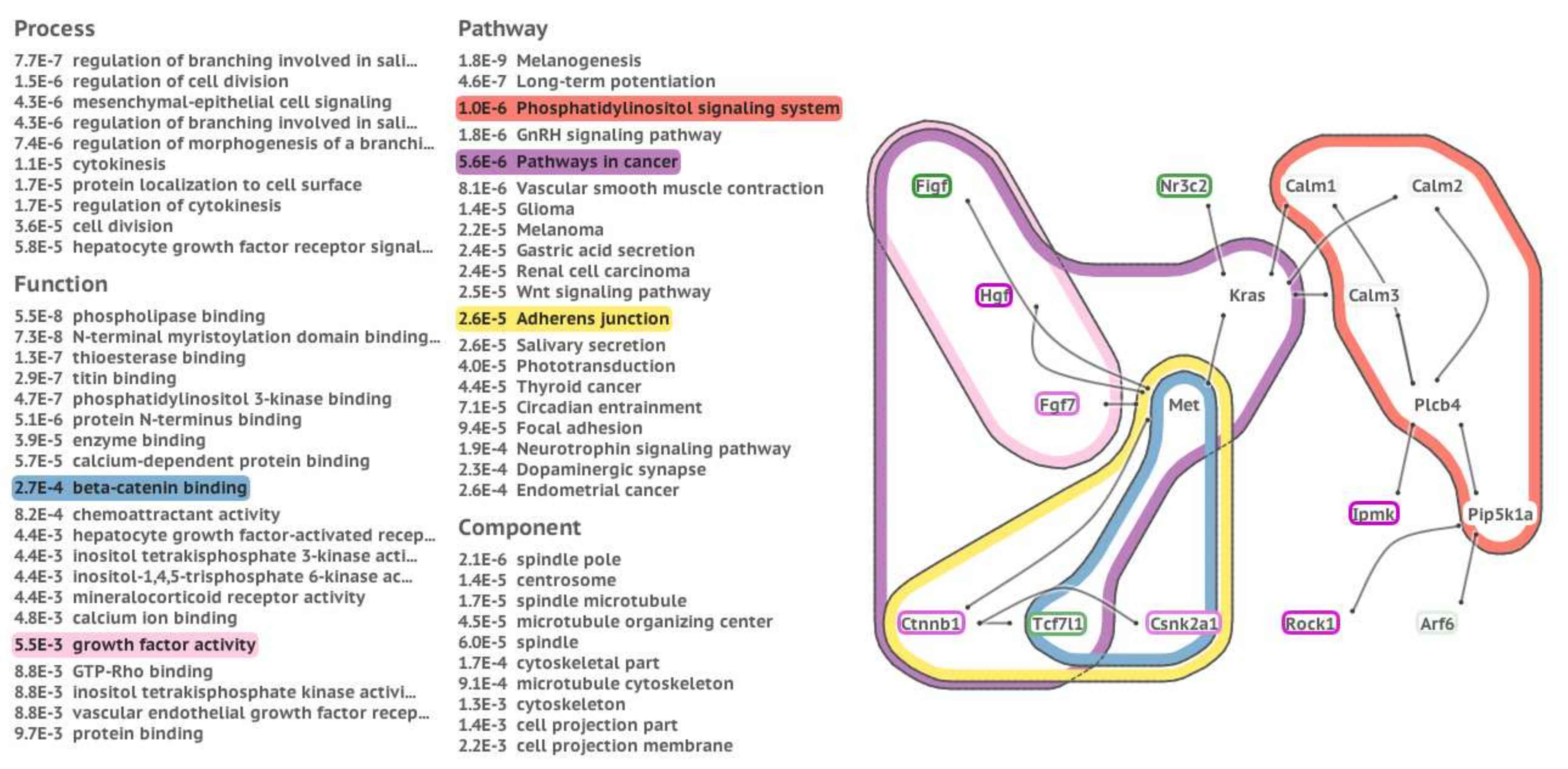}}
  \caption{\csentence{Overview of \examine.} (a)~In the \examine control panel the user can choose which
    categories to visualize. (b)~\examine presents an overview of available sets at the left
    and the annotated module visualization at the right. Here, the GO terms
\emph{beta-catenin binding} and \emph{growth factor activity}, and KEGG pathways
\emph{Phosphatidylinositol signaling}, \emph{Pathways in cancer}, and \emph{Adherens junction} are
selected for inclusion as sets in the network view. Sets are assigned unique colors, where
the set overview serves as a legend. Gene expression (log fold change) is visualized via superimposed
rectangles on the text labels, integrating the attribute color mapping functionalities of Cytoscape.}
  \label{f:tool}
\end{figure}

We implemented the described technique in a Cytoscape app with a design
that emphasizes simplicity of interaction and visual presentation. The available
sets are listed in the \emph{set overview} on the left, where the user may
select sets for inclusion in the annotated \emph{network visualization} to the
right---see Fig.~\ref{f:tool}.

We use Java's default graphics API
to render the presented visualizations. Geometric operations on the contours,
such as dilations and erosions, are performed via Java Topology Suite~\cite{jts03}.
All described functionalities can be used at interactive
speeds for networks up to dozens of nodes, edges, and active sets, including the
laying out of the network with the \emph{RSOM} training algorithm.

\begin{figure}[hbtp]
  \centering
  \subfloat[]{\includegraphics[width=0.48\textwidth]{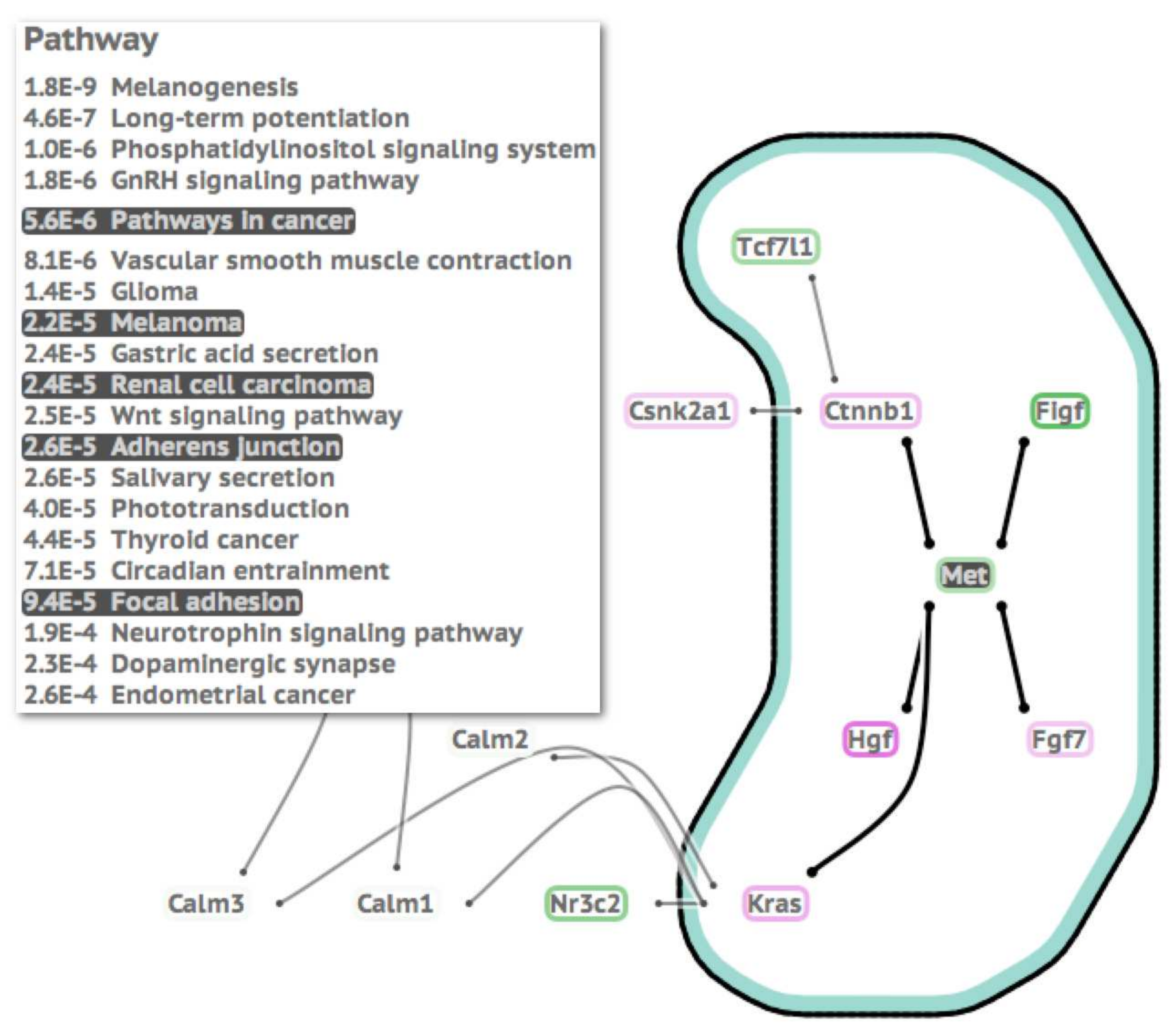}}\hfill
  \subfloat[]{\includegraphics[width=0.48\textwidth]{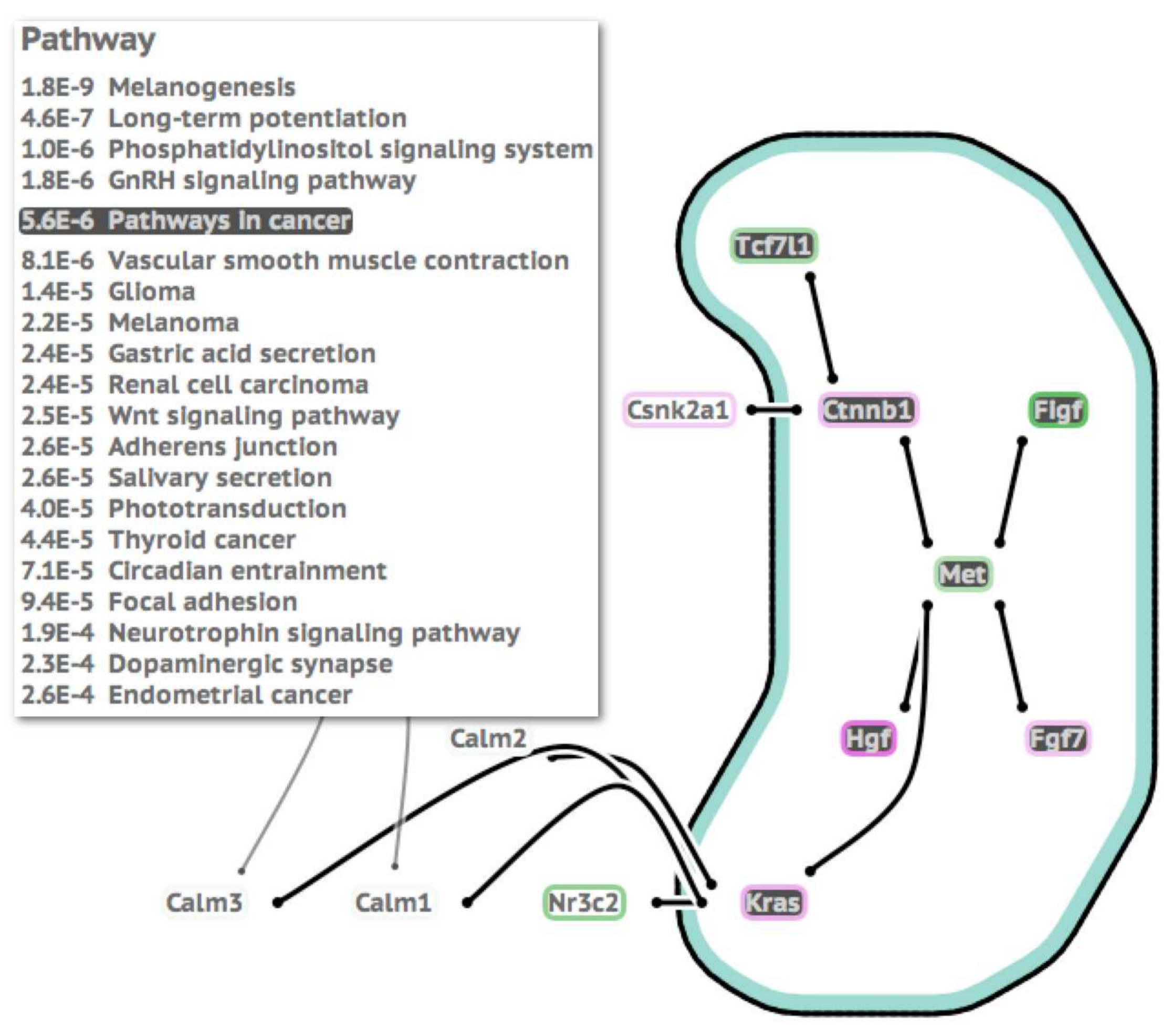}}

  \caption{\csentence{Item highlighting.} 
(a) Hovered protein (Met) with emphasized interaction links to its
neighbors, and sets (KEGG pathways) that contain this protein are highlighted as well.
(b) Hovered set (Pathways in cancer) with emphasized member proteins, interactions, and contour.}
  \label{f:highlight}
\end{figure}

\medskip\noindent\textbf{Interaction.}
Interactions consist of simple mouse actions (see the video in the
  Supplemental Material).
The inclusion of a set in the network visualization is toggled via
the set's label in the set overview or its contour in the network visualization.
Additional information about a set or node may be obtained via a hyperlink
to a web page provided in the input data.
This approach keeps the tool flexible, i.e., the tool itself does not have to be
altered every time a new kind of set or node from a different database is
loaded.
The links of a node are emphasized when it is hovered over (see Fig.~\ref{f:highlight}(a))
such that its direct neighborhood can be discerned from its surroundings. Moreover, sets that contain the hovered
node are highlighted as well. Vice versa, the contours of a set are emphasized and
its comprising nodes are highlighted when it is hovered over (see Fig.~\ref{f:highlight}(b)).
This provides immediate feedback to the user about node-set relations
without having to select a set and consequently changing the layout
of the network visualization.

The user can adjust the dominance of a set by hovering over either the set's
label in the set overview or set's contour in the network visualization, and
subsequently spinning the mouse wheel. This enables the user to give a set a
central role in the layout (see Fig.~\ref{f:dominance}(a)) or to
remove any of its influence (see Fig.~\ref{f:dominance}(b)).
All changes to the visualization caused by interaction are animated.
Colors and positions of items are altered gradually.
Link layout changes are animated by the interpolation of their control points,
while contour layouts are handled by fading out the old contour and fading in the
new contour. The use of layout preservation, as described previously, in
combination with animations helps to preserve the user's mental map.

\medskip\noindent\textbf{Color.}
We assign selected unique, distinguishable colors derived from
Color~Brewer palettes~\cite{brewer} in 
a cyclic manner so as to avoid the consecutive assignment
of the same color. In addition, we avoid large differences in contrast. For
example, we color text and set outlines
dark gray instead of black in order to reduce their visual dominance.
We only use intense black colors when items are hovered over or highlighted
such that they attract attention, as shown in Fig.~\ref{f:highlight}. Moreover,
we emphasize labels of selected sets (in the set overview) with a more intense
black color, retaining their legibility in a colored surrounding.
Node labels have a white background to make sure that their text is
legible when drawn on top of a set ribbon with a dark color. Likewise,
links have halos that make them easier to distinguish and their
intersections more pronounced.

\medskip\noindent\textbf{Cytoscape integration.}
\examine is tightly integrated into Cytoscape. We use Cytoscape's group
functionality to represent sets and rely on the table import functionality for
importing both the set and node annotations. We allow the user to group sets
into different categories.
We use the node fill color map attribute to color the node labels in \examine.
The user can invoke \examine on the currently selected nodes via the \examine
control panel. There the user can select which categories to show as well as the
number of sets per category. In addition, the user can specify that the
Cytoscape selection should be updated to match the union or intersection of the
selected sets in \examine---see Fig.~\ref{f:tool}.

\section*{Results and Discussion}
We will demonstrate how a domain expert can use \examine by working
out a case study in which we re-analyze a data set that some of
the co-authors have studied extensively. 


\subsection*{Case study of US28-mediated signaling in Human Cytomegalovirus}
\label{s:results}

We demonstrate our approach by a case study
involving the Human Cytomegalovirus (HCMV), a specific type of herpes virus with
a high incidence rate of 60\% among humans \cite{Gandhi04}.
Infection with HCMV in immune-competent individuals usually does not result in any
symptoms. However, in immune-compromised patients the virus is correlated with
pathologies such as hepatitis and retinitis \cite{Soderberg06}.
Although not considered an oncogenic virus, HCMV components 
have been detected in various tumors, giving rise to the hypothesis the virus
may act as an oncomodulatory factor in onset and development of cancer
\cite{Cobbs02,Harkins02,Cinatl04}.
The high levels of latent infection and the potential role in pathology of the
HCMV virus partly are associated with the HCMV-encoded G protein-coupled
receptors (GPCRs). Of these viral GPCRs, US28 is most studied and is
characterized as a chemokine
sink \cite{RandolphHabecker02}. Moreover, US28 is a
promiscuous, constitutively active viral GPCR, which hijacks the host cell’s
signaling pathways and stimulates proliferative, anti-apoptotic responses
interfering with natural programmed cell
death~\cite{Casarosa01,Maussang06,Maussang09,Slinger10,Langemeijer12}.
In previous studies, we performed transcriptome analysis to evaluate
US28-mediated pathways involved in pathological signaling. So far, a
gene-centric approach was used to identify differentially regulated genes
involved in HCMV-mediated pathology \cite{Maussang09,Slinger10}.

\begin{figure}[hbtp]
  \centering
  \subfloat[]{\label{f:venn}\includegraphics[width=0.4\textwidth]{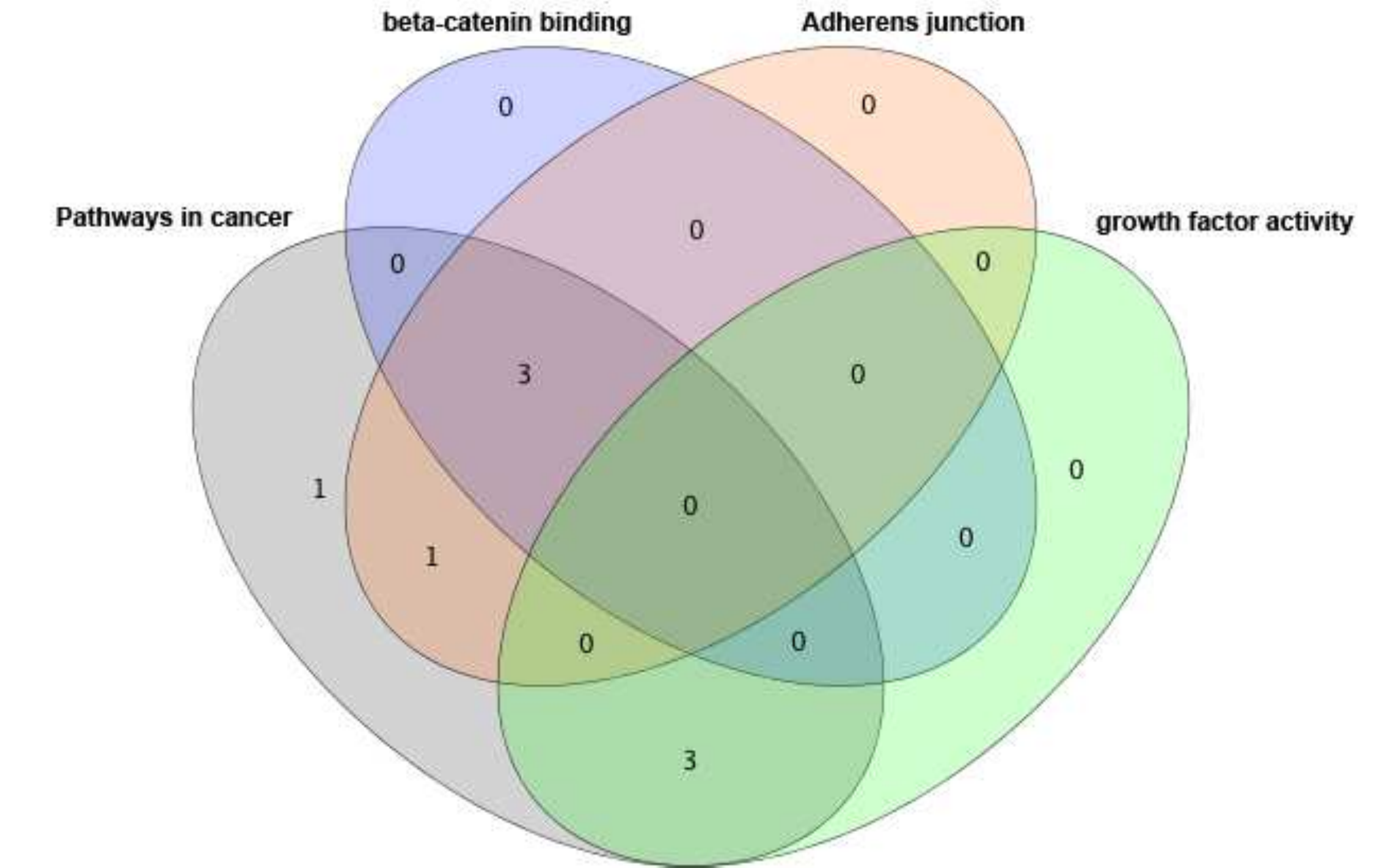}}
  \hfill
  \subfloat[]{\label{f:euler}\includegraphics[width=0.3\textwidth]{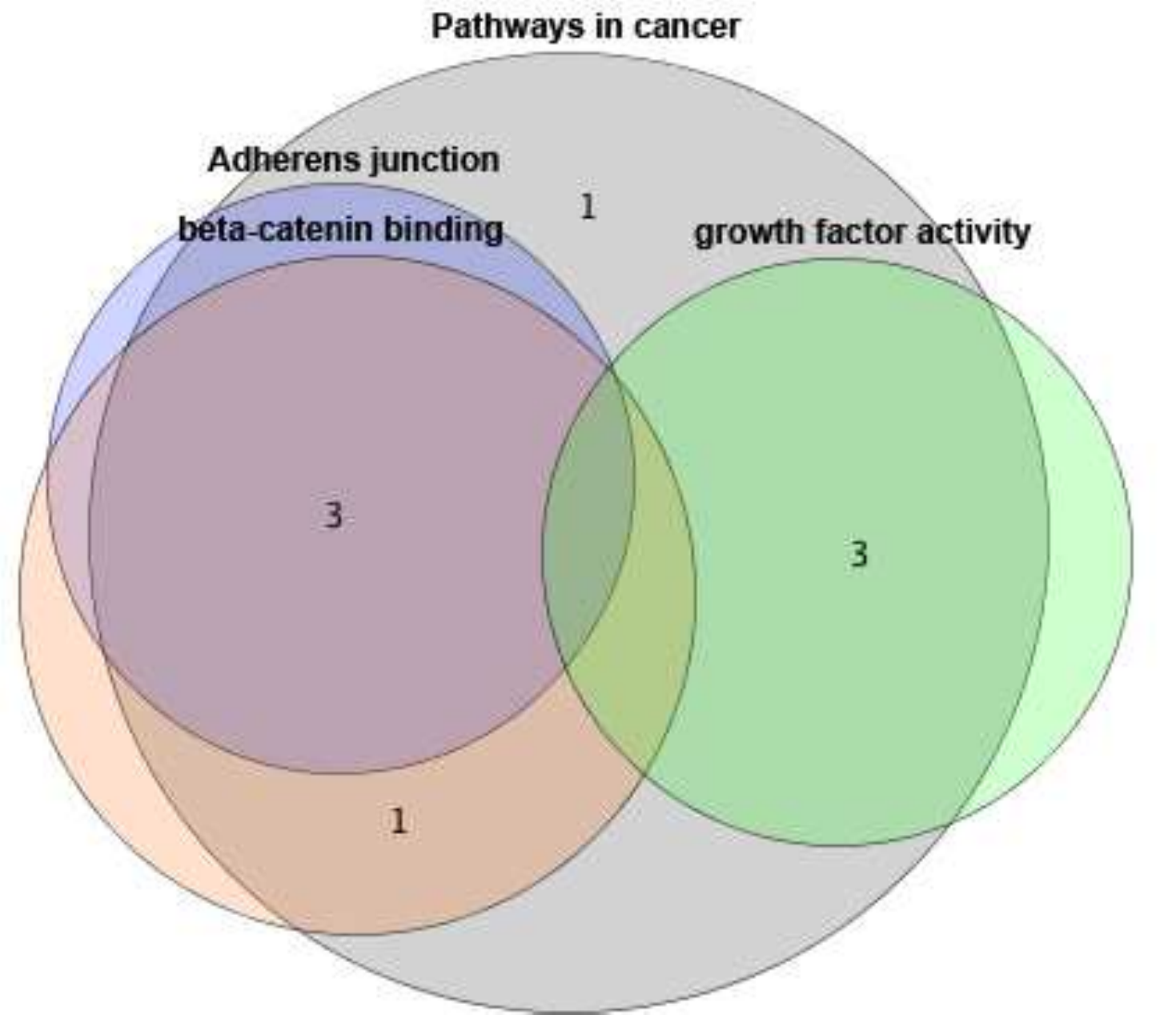}}
  \hfill
  \subfloat[]{\label{f:bubble_sets} \includegraphics[width=0.25\textwidth]{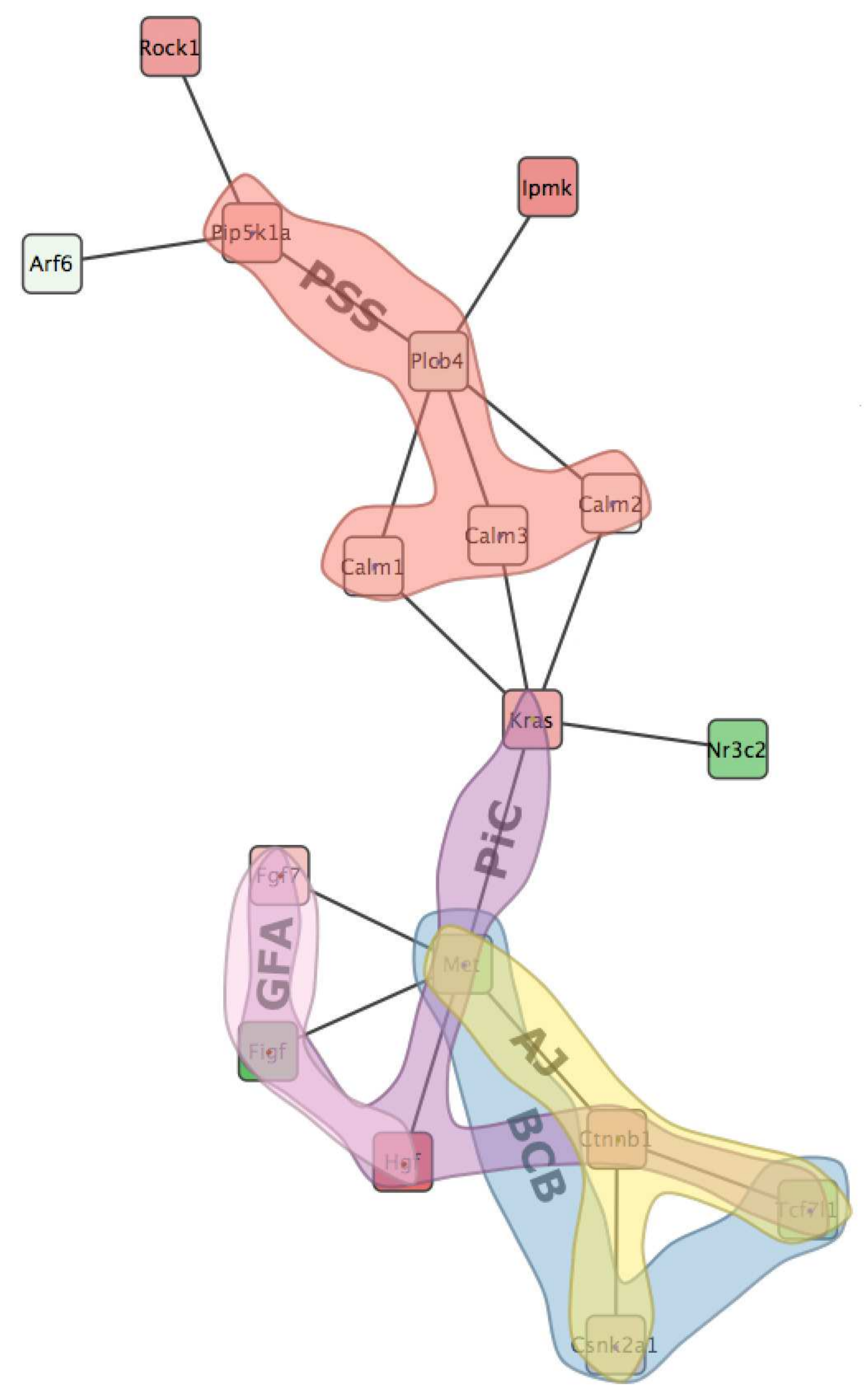}}
  \caption{\csentence{Comparison.} Annotated module visualization using Cytoscape's Venn and Euler
  diagram app: (a) Venn diagram and (b) Euler diagram. The number of displayed
sets is limited to four and no network structure is shown. (c) Module
laid out by one of Cytoscape's built-in force-directed layout
algorithms and BubbleSets superimposed on the network (same color
scheme as in Fig.~\protect\ref{f:tool}(b)). Note that it is not
immediately apparent that the nodes in the $\beta$-catenin set (blue) form a
subset of Adherens junction (yellow), because the BubbleSet approach
applies no explicit nesting of subsets.}
  \label{f:venn_euler_tool}
\end{figure}

In order to further identify the potential signaling properties associated with
US28, we used \examine to analyze the same data overlaid on the KEGG mouse
network~\cite{Kanehisa:2000jn}. The overlaid input network consisted of 3863 nodes
and 29293 edges. We computed $p$-values reflecting whether a gene is
significantly deregulated using RMA~\cite{gautier04} and 
LIMMA~\cite{Smyth05}. By running Heinz
\cite{Dittrich08}, a tool for identifying deregulated modules, we obtained a
deregulated module of 17 proteins. We performed an enrichment analysis to
determine enriched GO-terms and KEGG pathways. Fig.~\ref{f:tool} shows the
identified module---the steps taken for obtaining this visualization are given in the Supplemental Text. 
Parts of the same module are visualized using Cytoscape's Venn and
Euler diagram app in Fig.~\ref{f:venn} and \ref{f:euler}. The network structure is not shown in
this visualization and the number of displayed sets is limited to
four. Fig.~\ref{f:venn_euler_tool}(c) shows the module laid out by one of
the built-in force-directed layout algorithms of Cytoscape. All five
sets (see Fig.~\ref{f:tool} for the color scheme) are shown as
BubbleSets. The structure of the sets is not easy to understand: the
nodes belonging to the $\beta$-catenin binding set (blue shape) form a subset of
the Adherens junction set (yellow shape). Yet, BubbleSets does not
show these two shapes as nested structures.

We start by obtaining a global overview of the module by considering the
significantly enriched KEGG pathways. In line with the proposed oncomodulatory
role of US28, the module
shows significant enrichment for `Pathways in cancer' (purple
contour in Fig.~\ref{f:tool}).
In addition, the identified module is significantly enriched for
`Phosphatidylinositol signaling' (red contour), which corresponds to previous
work linking US28 to PLC-mediated calcium responses \cite{Casarosa01,Minisini03}.

Next, we focused on the `Pathways in cancer' submodule. By hovering over Tcf7l1,
Csnk2a1 and Met, we could see that these proteins are enriched for the
GO-term `beta-catenin binding'. By looking at the expression color coding, we
can see that $\beta$-catenin (Ctnnb1) and Csnk2a1 are up-regulated, whereas Tcf7l1 is
down-regulated. Normally, up-regulation of $\beta$-catenin
expression would increase output of the pathway as measured by Tcf7l1 expression
levels. However, up-regulation of Csnk2a1 negatively affects
$\beta$-catenin-mediated output, which is reflected by Tcf7l1 being down-regulated in
the module. Recently, we have studied
US28-mediated activation of $\beta$-catenin \cite{Langemeijer12}. In that
study, involvement of the upstream WNT/Frizzled pathway components activating
$\beta$-catenin was ruled out and an alternative mechanism involving Rock1,
which is also part of our module, was proposed. In the following we propose
another mechanism for US28-mediated activation of $\beta$-catenin.

\begin{figure}[hbtp]
  \centering
  \includegraphics[width=0.9\textwidth]{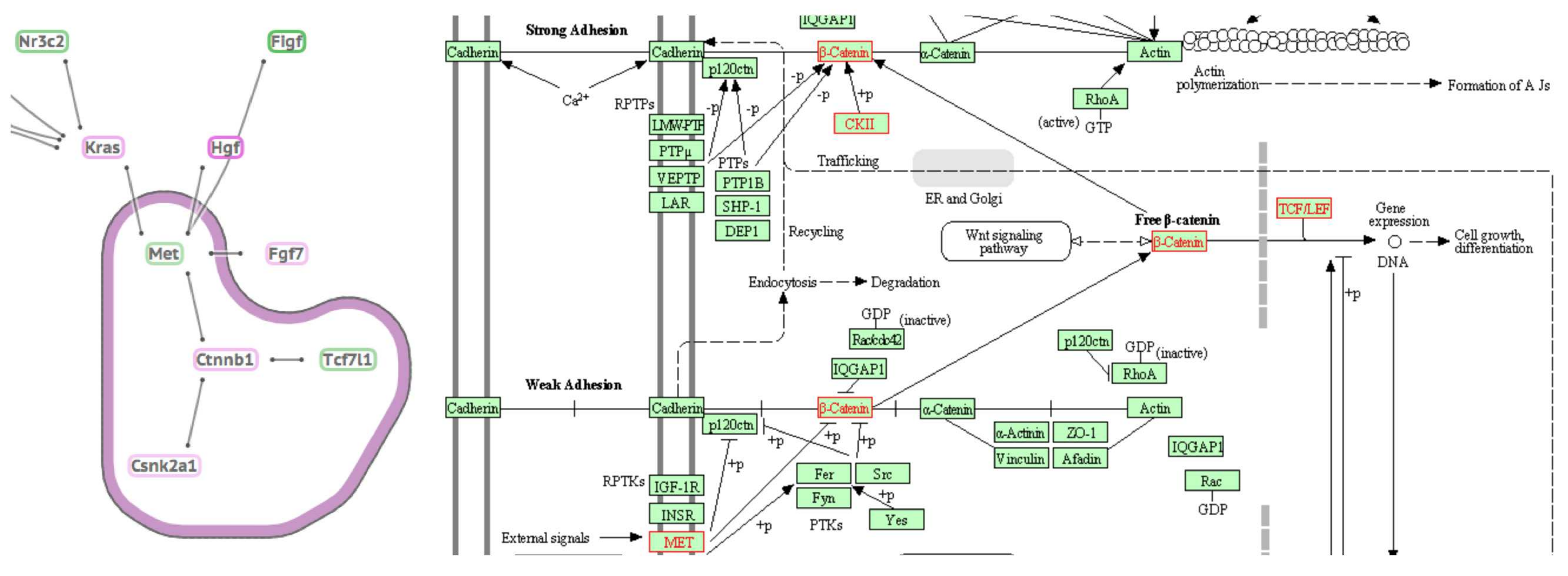}
  \caption{\csentence{Connection between Met and $\beta$-catenin.} Proteins that are associated to the selected \emph{Adherens junction} at the left
and corresponding KEGG pathway information at the right, where reactions catalyzed by module proteins are marked in red.}
  \label{f:pathway}
\end{figure}

By hovering over Fgf7, Hgf and Figf, we could see that these proteins are
enriched for the GO-term `growth factor activity' (pink contour). These
proteins are connected to $\beta$-catenin via Met. By requesting additional
information, we can see that Met is a receptor tyrosine kinase. Using
the plugin, a connection between Met and $\beta$-catenin is found via the KEGG
pathway `Adherens junction' (Fig.~\ref{f:pathway}). Indeed, alternative mechanisms in the activation of
$\beta$-catenin signaling involving release of $\beta$-catenin from cell-cell
adherence junctions have been described (e.g.\ \cite{Zhurinsky00}). 
As US28 was also found to
mediate cell migration \cite{Streblow99,Streblow03}, loss of cell-cell contacts
may explain increased levels of $\beta$-catenin signaling as observed in
US28-expressing cells. When the observed increase of Hgf gene expression is
reflected in increased excretion of this growth factor in US28-expressing cells,
this may give rise to the following hypothesis explaining the enhanced TCF-LEF
reporter gene activation in US28-expressing cells. Increased Hgf may activate the
Met receptor tyrosine kinase in an auto- or paracrine fashion. The Met tyrosine
kinase has been shown to mediate release of $\beta$-catenin from adherence
junctions associated with the phosphorylation of $\beta$-catenin
on tyrosine 654 \cite{Herynk03,Purcell11}.
This phosphorylation may result in increased levels of $\beta$-catenin in the
cytoplasm which could explain enhanced TCF-LEF activation. We are currently
experimentally validating this hypothesis.

\subsection*{Discussion}
\label{s:discussion}
The case study demonstrates that our set-oriented visualization
approach is well suited for analyzing protein modules enriched with
gene expression data, Gene Ontology annotations, and KEGG pathway
information.  The visualization itself uses familiar visual concepts,
such as node-link diagrams and contours drawn around items that belong
to the same set, which makes interpretation intuitive. Our layout
algorithm, based on SOMs, constructs an integrated layout of both network
topology and set system topology. It allows for emphasizing either
network topology or set system topology via a single parameter that is
adjustable interactively by the user. Existing approaches do not offer
this flexibility, and will either optimize the layout of the set
relations at the expense of the network topology or emphasize network
topology to the detriment of the set system.

Other types of (network) visualization, such as adjacency matrices and
enrichment tables, could also be used to integrate additional set
data.  However, these visualizations are not as intuitive and, more
importantly, they are less effective in conveying the topology of the
network and set system.

Scalability is a limitation of our approach: if the network is very
large and if there are many sets, it is not possible to construct a
comprehensive layout, which makes visual analysis ineffective. This is a natural limitation of any
visualization approach based on node-link diagrams and set
contours. Our technique relies on a focus and context approach, in
which the network and set system should be pruned (computationally)
down to the most relevant or significant components first. The resulting
smaller network module(s) and set system can then be visually explored
with our approach, and the domain expert can focus on the relevant
details only to understand the underlying biology.

\section*{Conclusions}
\label{s:conclusion}
We have proposed a set-oriented visual analysis approach for annotated network
modules, where sets are shown as contours on top of a network node-link layout. 
We have implemented this approach
in the Cytoscape app \examine that enables the interactive exploration of
annotated network modules. 
Subsequently, we have used the tool to re-analyze a data set that
some of the co-authors have studied extensively. As a result of using the plugin,
we have been able to formulate a new hypothesis about deregulated signaling of
$\beta$-catenin by viral receptor proteins. This new hypothesis is currently
being verified experimentally. 

\section*{Availability and Requirements}

\textbf{Project name:} \examine\newline
\textbf{Project homepage:}
\url{http://apps.cytoscape.org/apps/examine}\newline
\textbf{Operating system(s):} all\newline
\textbf{Programming language:} Java\newline
\textbf{Other requirements:} Cytoscape 3.x\newline
\textbf{License:} GPL2\newline
\textbf{Any restrictions to use by non-academics:} None



\begin{backmatter}

\section*{Competing interests}
  The authors declare that they have no competing interests.

\section*{Author's contributions}
KD, MEK, MAW and GWK conceived the visual analysis technique. KD and
MEK implemented \examine. CIB, MEK, GWK, MS and MJS applied it to the
US28 case study. KD, MEK, MAW and GWK drafted the manuscript. All authors read and approved the final manuscript.   


\bibliographystyle{bmc-mathphys} 
\bibliography{eXamine}      

\end{backmatter}
\end{document}